\documentclass{article}

\usepackage{arxiv}

\usepackage[utf8]{inputenc} % allow utf-8 input
\usepackage[T1]{fontenc}    % use 8-bit T1 fonts
\usepackage{hyperref}       % hyperlinks
\usepackage{url}            % simple URL typesetting
\usepackage{booktabs}       % professional-quality tables
\usepackage{amsfonts}       % blackboard math symbols
\usepackage{nicefrac}       % compact symbols for 1/2, etc.
\usepackage{microtype}      % microtypography
\usepackage{lipsum}
\usepackage{fancyhdr}       % header
\usepackage{graphicx}       % graphics
\graphicspath{{media/}}     % organize your images and other figures under media/ folder
\usepackage{algorithm}
\usepackage{algorithmic}

\usepackage{latexsym}

\usepackage{subcaption}
\usepackage{mwe}

\title{Latent Sub-structural Resilience Mechanisms in Temporal Human Mobility Networks during Urban Flooding
\thanks{\textbf{To whom correspondence should be addressed. E-mail: amostafavi@civil.tamu.edu}}: 
}

\author{
  Akhil Anil Rajput, Ali Mostafavi \\
  Zachry Department of Civil Engineering  \\
  Texas A\&M University \\
  College Station, TX, USA\\
  \texttt{akhil.rajput@tamu.edu, amostafavi@civil.tamu.edu} \\
}

\begin{document}
\maketitle

\begin{abstract}
In studying resilience in temporal human networks,  relying solely on global network measures would be inadequate; latent sub-structural network mechanisms need to be examined to determine the extent of impact and recovery of these networks during perturbations, such as urban flooding. In this study, we utilized high-resolution aggregated location-based data to construct temporal human mobility networks in Harris County, Texas (Houston metropolitan area) in the context of the 2017 Hurricane Harvey. Using the constructed temporal network models, we examined characteristics such as motif distribution, motif persistence and temporal stability, and motif attributes to reveal latent sub-structural mechanisms related to the resilience of human mobility networks during disaster-induced perturbations. The results show that urban flood impacts persists in human mobility networks at the sub-structure level for several weeks. The impact extent and recovery duration is heterogeneous across different network types. Also, perturbation impacts persist at the sub-structure level while global topological network properties might indicate the network has recovered. The patterns of impact and recovery at the global-network scale is influenced by more abundant motifs; however, less abundant motifs (which are shown to have more stability during steady state) experience greater sustained impacts and take longer to recover. The findings highlight the importance of examining the microstructures and their dynamic processes and attributes in understanding the resilience of temporal human mobility networks (and other temporal networks), it is essential to examine the microstructures and their perturbation impacts and recovery. The findings can also provide disaster managers, public officials, and transportation planners with insights to better evaluate impacts and to monitor recovery in affected communities based on the patterns of impact and recovery in human mobility networks at both sub-structure and global-network levels.
\end{abstract}

% keywords can be removed

\section*{Introduction}
Characterizing network resilience has received attention in natural ~\cite{gao_barzel_barabasi_2016}, social ~\cite{Janssen2006TowardAN, RAJPUT2020101622}, physical ~\cite{Bhatia2015NetworkSB, Liu2021LeveragingNT}, and engineered systems ~\cite{Hosseini2014ResilienceMA}. In particular, to reduce the impacts of disasters, a growing body of literature has examined the resilience of various socio-spatial and physical networks during disasters. In particular, over the past few years, human mobility networks have been  studied in the context of crises, such as floods and pandemic, to advance the understanding of impacts and recovery of communities ~\cite{jiang2017activity, BARBOSA20181, yabe2020non, changruenngam2020individual, Spyratos2019, Zhao2021}. There have been multiple recent research streams examining mobility networks that focus on examining evacuation patterns, movement fluctuations, resilience of communities to disasters  \cite{podesta_coleman_esmalian_yuan_mostafavi_2021}, and equitable resilience of disasters in communities \cite{coleman2021human} in the context of hurricanes and flood events. A common idea across all these studies is that analyzing changes in the structural attributes of human mobility networks can reveal insights on mechanisms of failure and recovery to allow stakeholders and researchers tools to better evaluate and quantify the impacts and recovery duration, and to draw insights to improve resilience for future events. Despite the growing number of studies on human mobility networks and their resilience characteristics during crises, the majority of the existing work focuses on topological measures at the network level; limited attention has been paid to resilience characteristics at the sub-structure/subgraph level.

Subgraphs (motifs) and their characteristics play an important role in topological and dynamical properties of networks, in particular, temporal networks. However, the current state of the art is rather limited in terms of characterization of perturbation impacts and recovery of network motifs. Hence, examining network motifs and their characteristics could move us closer to a more complete characterization and understanding of network resilience and recovery in many real-world networks, such as spatial-temporal urban networks.

A motif, defined as subgraph, consists of a few nodes that are embedded in a larger graph ~\cite{Prill2005}. Motif substructures are deemed as the building blocks of most networks ~\cite{Milo2011}. However, a majority of the existing literature related to network resilience and recovery focuses on non-temporal networks in which the effects of perturbations are examined based on fluctuations in the global network properties (such as giant component size), and network recovery is quantified based on the time it takes for global network properties to be restored to their steady state level. This approach to characterization of network resilience and recovery is particularly problematic in temporal networks where local topological and dynamical properties that vary with time may not assessed with traditional measures. This limitation has led to an incomplete understanding and characterization of network resilience and recovery. 

Recently, limited studies related to characterizing urban mobility networks based on their motifs have primarily focused on understanding structurally and spatially heterogeneous patterns of urban mobility networks \cite{Cao2021}, finding differences between telecommuters and commuters \cite{Su2021}, modeling of human activity and mobility \cite{Jiang2016, jiang2017activity},  and urban traffic speed prediction \cite{Zhang2020}. A number of other studies have examined motifs in transportation networks (such as road networks and airline networks). ~\cite{Iovanovici2019} used motifs on transportation networks to better understand their topology. \cite{Jin2019}, in  calculating motifs for airline networks found that adjusting the number of proper network motifs is useful to optimize the overall structure of airline networks for profitable air transport.

Subgraphs (motifs) and their characteristics play an important role in topological and dynamical properties of networks (and temporal networks in particular). The current state of the art, however, is rather limited in terms of characterization of perturbation impacts and recovery of network motifs. Hence, examining network motifs and their characteristics could move us closer to a more complete characterization and understanding of network resilience and recovery in many real world networks, such as spatial-temporal urban networks; however, very limited studies have attempted to characterize network resilience in disasters based on the dynamic properties of motifs. \cite{Dey2019} studied network resilience and reliability under various types of intentional attacks using motifs on electricity transmission networks of four European countries. The findings of the study highlight the importance of characterizing the dynamic properties of motifs in assessing network resilience. In particular, we hypothesize that, in temporal networks, characterization of network resilience based on motif properties may reveal extents and patterns of impacts and recovery at the substructure level different from the patterns observable at the network level. For example, In studying brain networks, \cite{Duclos2021} found that motifs show re-organization during loss and recovery of consciousness but global network properties are not able to consistently distinguish between responsive and unresponsive states. A similar phenomenon might also exist in human mobility networks during flood-induced perturbations. To test this hypothesis, we utilized high-resolution location-based data to construct and analyze human mobility networks in Harris County, Texas in the context of the 2017 Hurricane Harvey and its flooding to uncover dynamic properties of motifs during perturbation and early recovery and draw insights regarding differences in resilience patterns at a subgraph level versus a global-network level.

Hence, the main idea motivating this study is that the impacts of perturbations on the characteristics of network subgraphs (motifs) and their characteristics could vary from what could be observable using the global network properties; hence, by examining perturbation impacts and recovery at the subgraph level, provides a more complete understanding of network resilience and recovery in temporal networks. To this end, we evaluate the effect of disasters on network sub-structures (motifs) in the mobility networks using census tract-level mobility data for Harris County at the time of Hurricane Harvey. We start by creating a spatio-temporal representation of human mobility using network graphs. We then investigate characteristics exhibited by sub-structures in the network. In particular, we compute motif concentration; motif persistence, i.e., analogous to persistence homology; and motif conversion trends to assess pathways to recovery. We also calculate motif attributes such as travel volume and travel distance attributed by different motif types. We compare the results obtained from analyzing network sub-structures with temporal fluctuations in global network properties, such as network diameter and modularity, to address of the primary question of whether global networks can adequately capture hidden mechanisms of insatiability in mobility networks. Accordingly, we address the following research questions:
\begin{enumerate}
    \item To what extent the distribution, stability, and attributes of different motifs vary during steady state?
    \item What is the extent of flood-induced perturbations on motif characteristics, such as distribution, stability, attributes and their recovery?
    \item What differences exist among patterns of network recovery at global-network scale versus at the motif-distribution and attribute level?
\end{enumerate}

The results and findings move us closer to a more complete understanding of resilience characteristics in human mobility networks during disasters. The results also provide useful insights for transportation planners, disaster managers, and public officials to better evaluate and monitor impacts and recovery in human mobility networks during crises. Also, The outcomes from this study could provide new and valuable insights for understanding resilience in other networks, such as transport, ecological, and biological networks based on detecting sub-structure instability before they are manifested in global network properties. 

\section*{Data and Methods}
\subsection*{Data description and network generation}

\begin{figure}[tbhp]
\centering
\includegraphics[width=.8\linewidth]{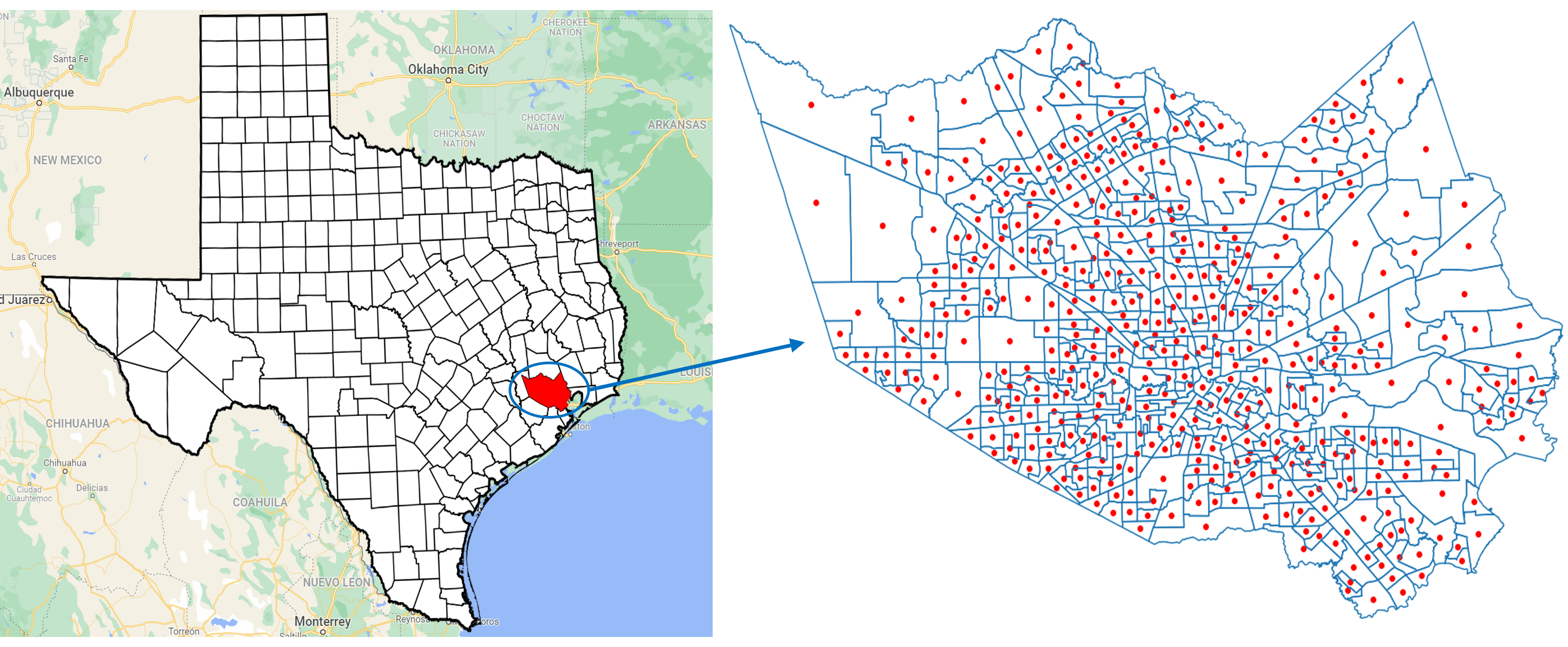}
\caption{Spatial distribution of modified census tracts that are used as spatial units to aggregate data and construct the human mobility network for Harris County. Red nodes represent the centroid of each modified CT and also the nodes in the O-D network.}
\label{fig:HC_CTs}
\end{figure}

We used mobility data provided by StreetLight, a commercial platform which aggregates and anonymizes movements between spatial areas using smartphones as sensors to measure vehicular, transit, bicycle, or foot traffic. StreetLight Data provides origin-destination (O-D) analysis data. Our analysis aggregated anonymized data from cell phones and GPS devices to create travel metrics, such as duration and distance (StreetLight 2021a). These data incorporate the trips using different modes of transportation, including personal cars and public transit. By analyzing more than 40 billion anonymized location records across the United States in a month and enriching the analysis with other sources, such as digital road network and parcel data (StreetLight 2021b), StreetLight Data is capable of reaching approximately 23\% penetration rate (InSight 2018), covering distinct census divisions in North America’s road network. Thus, the data provide a proper sample of human mobility in Harris County for examination of grocery store access disparity in the face of Hurricane Harvey. We used census tracts (CTs) as spatial areas for data aggregation and constructing the spatial human mobility network. Harris County has more than 800 CTs; due to restrictions from the data provider on the number of spatial units provided, we used modified census tract polygons for Harris County. Modified CTs were obtained by merging smaller CTs with their neighbors until we reached 500 CT polygons. Figure \ref{fig:HC_CTs} shows the spatial distribution of the obtained modified CTs. We used these modified CT polygons to get information on the aggregated number of trips between them at an hourly resolution from August 1, 2017 to September 30, 2017, that we later aggregated at daily period. 

To construct the human mobility network, we consider a graph $G=(V, E, \omega)$, where $V$ is the set of nodes, $E \subset V \times V$ is the set of edges, and  $\omega: V \times V \mapsto \mathbb{R}_{\ge 50}$ is an edge weight function such that each edge $e_{uv} \in E$ has a weight $\omega_{uv}$. The nodes represent the centroid of the modified CT polygons, the edges represent the presence of significant number of trips between the nodes, and $\omega$ is a mobility network weight function that represents the movement volume. The total number of nodes in $G$ is $n = |V|$. We use $G$ as an undirected network representation, i.e., for all $e_{uv} \in E$,  $e_{uv}\equiv e_{vu}$. The adjacency matrix of the network is defined by $A_{ij}=\omega_{ij}$, if $(i,j)\in E$, 0, otherwise. If $\omega_{ij}=1$, $G$ is called unweighted network.

Human mobility networks are temporal networks. The structure of a temporal network changes in time, which can be represented with a time-indexed graph $G_t=(V(t),E(t),\omega(t))$, where $V(t)$ is the set of nodes in the network at time $t$, $E(t) \subset V(t) \times V(t)$ is a set of edges in the network at time $t$, and  $\omega(t)$ is an edge weight function at time $t$. Here, $t$ is either discrete or continuous. For this study, we constructed a mobility network ($G_t$) for each day ($t$) between August 1, 2017 to September 30, 2017: $\mathbb{G} = \{{G}_{1}, \ldots, {G}_{\mathcal{T}}\}$, where $\mathcal{T}=61$. This timeframe includes the major events during Harvey, i.e., (1) August 23, 2017: Declaration of state of disaster for 30 Texas counties, (2) August 25, 2017: Harvey makes landfall in Harris County (Houston), and (3) August 29, 2017: Further flooding in Houston due to release of water from the Addicks and Barker reservoirs. We therefore define the Harvey period as the duration from August 25 to August 29, 2017, the timeframe during which Harris County suffered direct damage from the hurricane.

Temporal mobility networks generated from the StreetLight dataset directly had 500 nodes and roughly 30,000 edges in the nonperturbed state. Since our algorithm's run time depends on the size and complexity of the network, to reduce the complexity we consider only links that exhibit significant number of trips. We define significant trip number as an aggregated movement of more than 50 in a direction. Thus we added the edges only to the network that had movement of more than 50 in each direction. This approach did not significantly change the network structure as trips less than 50 accounted for only 3-5\% of the edges during nonimpact days. Therefore, including this threshold value will not change the results significantly.
To generate the final network, we added the links having movement in both directions to get an undirected origin-destination mobility network. We followed the same approach for network generation for all the days in the study period.

The number of trips between each modified census tract or the weight of the edges are expected to drop during the hurricane and flood period as movement may be restricted and usual commute is greatly impacted by road inundations. This may cause many edges to go under the threshold of 50 trips and thus they would not be considered as a part of the network. The lost edges imply that the movement between the CTs connected by  these edges are insignificant and hence in a perturbed or impacted state.

\subsection*{Network substructure, Motifs}

\begin{figure}%[tbhp]
\centering
\includegraphics[width=0.8\linewidth]{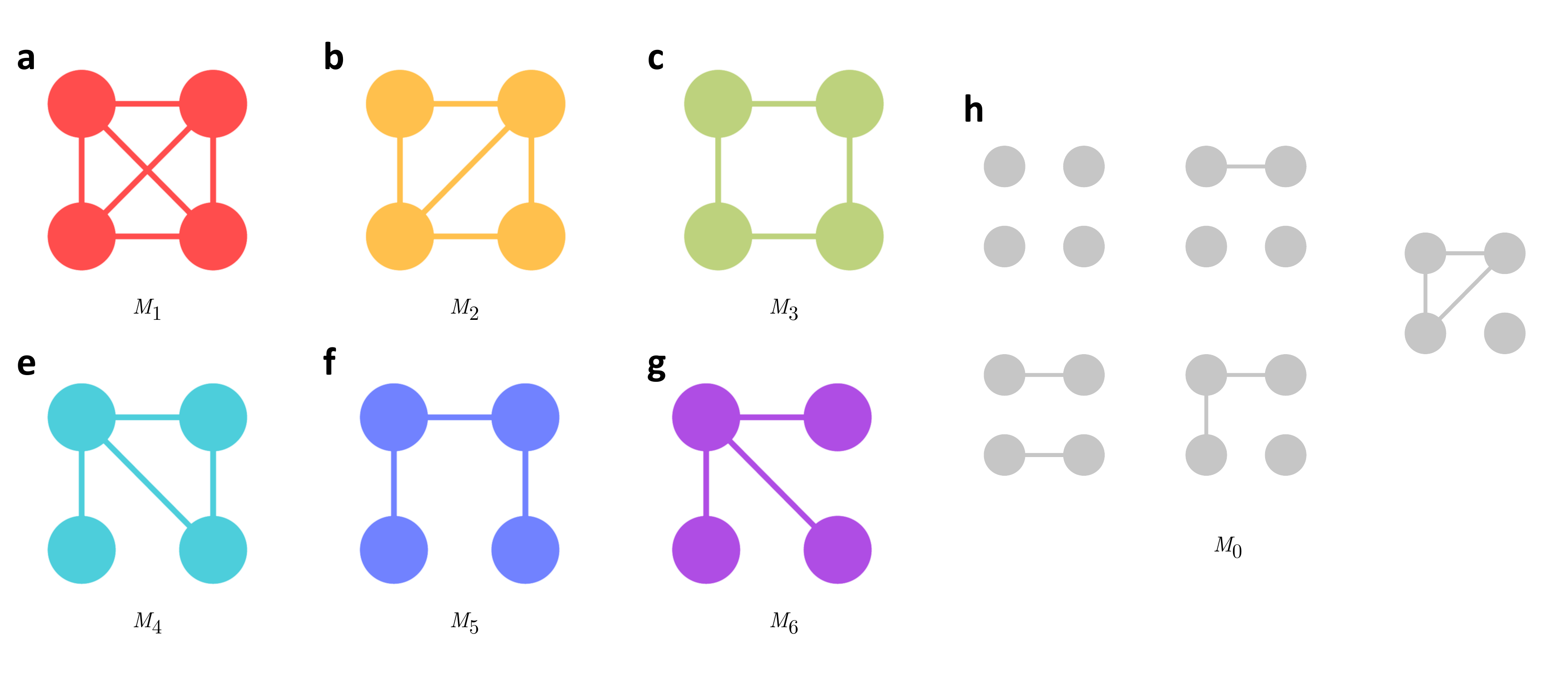}
\caption{Motif substructures used in this paper. (a) through(g) represent the six motifs that are the only way a four-node pair can be connected to form a connected subgraph. (h) symbolic motif type 0 that corresponds to disconnected four-node subgraphs.}
\label{fig:motifs}
\end{figure}

After constructing the mobility networks for different days, we examine the motifs in the networks. A motif, $G'=(V', E')$, is defined as a recurrent multi-node induced subgraph pattern in $G$, i.e., $V'\subseteq V$ and $E'\in E$, and $E'$ contains all edges $e_{uv}\in E$ such that $u, v\in V'$. For our analysis, we consider four-node motif structures as shown in Figure \ref{fig:motifs}. Global network properties focus primarily on global connectedness at the macroscopic level. Network motif analysis, on the other hand, captures micro structures and represents the local interaction pattern of the network. Motif types 1 and 2 represent densely connected subgraphs where almost all nodes are connected to each other. Therefore, they represent movement between areas that exhibit most interconnected movement patterns. Motif type 4 has a three-node cycle and an open edge. Motif type 3 and 5 may represent general commute patterns for work or other lifestyle patterns. Motif type 6 represents a hub and a spoke structure. Motif type 0 is a symbolic motif type that we use to identify four-node pairs where at least one node is disconnected from its subgraph and does not fall into any of the six motif types. 

\subsection*{Motif distribution}

From the daily temporal graphs related to human mobility, we calculated the motif count for each graph, $G_t$ representing each day from August 1, 2017 to September 30, 2017. Since the graph is dense with 500 nodes and nearly 30,000 edges, it is computationally very difficult to count every motif present in the network. To overcome this limitation, we chose 100,000 four-node pairs at random from the network and identified the motif type that they correspond to. This was under the assumption that 100,000 four-node pairs provide a good representation of the network. The motif count distribution does not change when we resample 10,000 or more nodes. Therefore, we can be confident that this approach provides near-similar distribution of actual motifs present in the network. 

In the first step of motif analysis, we examined how disasters or perturbation impact motif distribution in the network and determined the impact and recovery point in the network based on changes in motif distributions. We looked at the first two weeks of August as the baseline period and compared the motif distribution in terms of percentage of each type of motif in a day for the subsequent weeks. Baseline values were computed for each day of the week separately and were only compared to same days of the week. To account for differences in movement patterns over weekdays and weekends, we plotted results for weekdays and weekends separately.

We evaluated the changes in \textit{motif distribution} where the distribution ($D_i$) of an $n$-node motif of type $i$ can be defined as the ratio of its number of occurrences ($N_i$) to the total number of $n$-node motifs in the network, i.e., $D_i={N_i / \sum_{i} N_i}$, where $\sum_{i} N_i$ is the total number of $n$-node motifs. Also, to reduce the noise in the data, and remove the weekday-weekend effect, we looked at 7-day moving average for all days. We also looked at the absolute values for motif distribution for each of the days to evaluate the abundance of each motif type in the network to determine if the frequently occurring motifs exhibit higher temporal stability. 

\subsection*{Motif persistence and evolution}

In the next  step, we investigated the temporal stability of the motifs and introduced a new method of visualizing motif stability to identify stable motifs and their conversion trends. To evaluate the stability of motifs, we sampled around 900,000 unique four-node pairs from the graphs that contributed to at least one of the motif types for any of the days and tracked the motif type exhibited for each day. For example, if nodes 1,2,3, and 5 are one of the sampled pairs, then we track what motif type it corresponds to for each of the days. If it does not fall in any of the six motif categories shown in Figure~\ref{fig:motifs}, then we assign it a symbolic motif, motif type 0, which means that at least one of the nodes from the sampled pairs is isolated. 

We then visualized persistent diagrams for these motifs, analogous to persistent homology diagram representation~\cite{edelsbrunner2008persistent} where the birth event signifies occurrence of a particular motif on that day and death event corresponds to conversion to a different motif type or getting disconnected by losing one or more nodes from the four-node subgraph. For example, if we are looking at the stability of motif type 1, then the birth event signifies occurrence of motif type 1, and death event corresponds to conversion of this motif to some other motif type or getting disconnected (conversion to motif type 0). These persistent motif figures represent the life of each of the motif types and what conversion patterns they exhibit. In contrast to barcode approach ~\cite{ghrist2008barcodes}, where there is a single birth and death event, motifs can show multiple birth-death patterns. This is because motifs can convert back to their original motif type.  If a motif shows long life, then we can infer that particular motif type is persistent across all the days and thus more stable.

To better understand these conversion patterns and to identify the markers for recovery, we plotted the motif trends for each motif for all consecutive days. This is explained in more details in the result section.

\subsection*{Motif characteristics}

Along with structural characteristics of motifs in terms of structural changes, we also looked at how the intrinsic characteristics of these motifs change over time. To understand how motifs shape the characteristics of human mobility networks and the impact of disasters on these properties we looked at the change in average travel volume and the average distance metric for each motif type. Average travel volume for a motif is calculated by adding all the link weights for a motif subgraph. This metric represents the average travel volume that a motif is contributing to the entire network.

We first get a list of all four nodes sampled in our analysis to get motif distribution. We then split the four-node pairs into six categories based on their corresponding motif. For node pairs in each of these categories (node pairs belonging to a particular motif type) we get the motif subgraph and calculate the average of the edge weights (travel volume). We reported the median of these average values in each category. Absolute values of computed measure and change with respect to baseline are plotted for all days.

Average motif distance represents the distance range that a particular motif connects on average. To compute this metric, we first geocoded all the nodes in the network based on the location of the nodes that represent the centroids of CTs. Then, we calculated the distance represented by each of the links in the motif subgraph and calculated its average. For all the motifs of the same type, from the distribution of these average distance values, we reported the median value. Similar to average motif travel volume, we plotted the absolute and percent change for the median of the average motif distance value for all motif types for all days.

\subsection*{Global Network properties}

Global network properties such as giant component, network diameter, and modularity are studied to understand the impact of perturbation on networks and are comparatively straightforward to compute. To understand if temporal changes in the substructures of a network shows similar changes at the global scale, we looked at network properties such as average network diameter, modularity, density, average degree, and giant component size. Network diameter is a measure of the shortest distance between the two most distant nodes in the network. Networks with smaller diameters are more resilient, and those with larger diameter are sparse and have fewer redundant connections \cite{Zhang2015}. Modularity measures the degree to which a network's densely connected nodes can be decoupled into separate communities or modules \cite{Newman2006ModularityAC}. Increased modularity in a network safeguards the network against the spread of shocks, such as infectious diseases \cite{Kharrazi2020}. In the case of mobility networks, increased modularity would imply that the movements are more localized and happen mostly within different clusters which reflects that movements do not take place freely and thus increased modularity is not a desired trait.

Network density is an indicator of density connections in a network. It is computed by calculating the ratio of number of edges present in the network with the possible number of edges \cite{Scott1991SocialNA}. Average degree of a network gives an indicator of the average number of links present for each node in the network. Giant component size is an indicator of the number of nodes present in the largest connected component in a network. We calculated these global network measurements and their fluctuations during the study period to examine the extent of impacts and duration of recovery in human mobility networks for comparison with the patterns manifested at the sub-structure level.

\section*{Results}

\subsection*{Motif Distribution}

\begin{figure}
\centering
\includegraphics[width=0.7\linewidth]{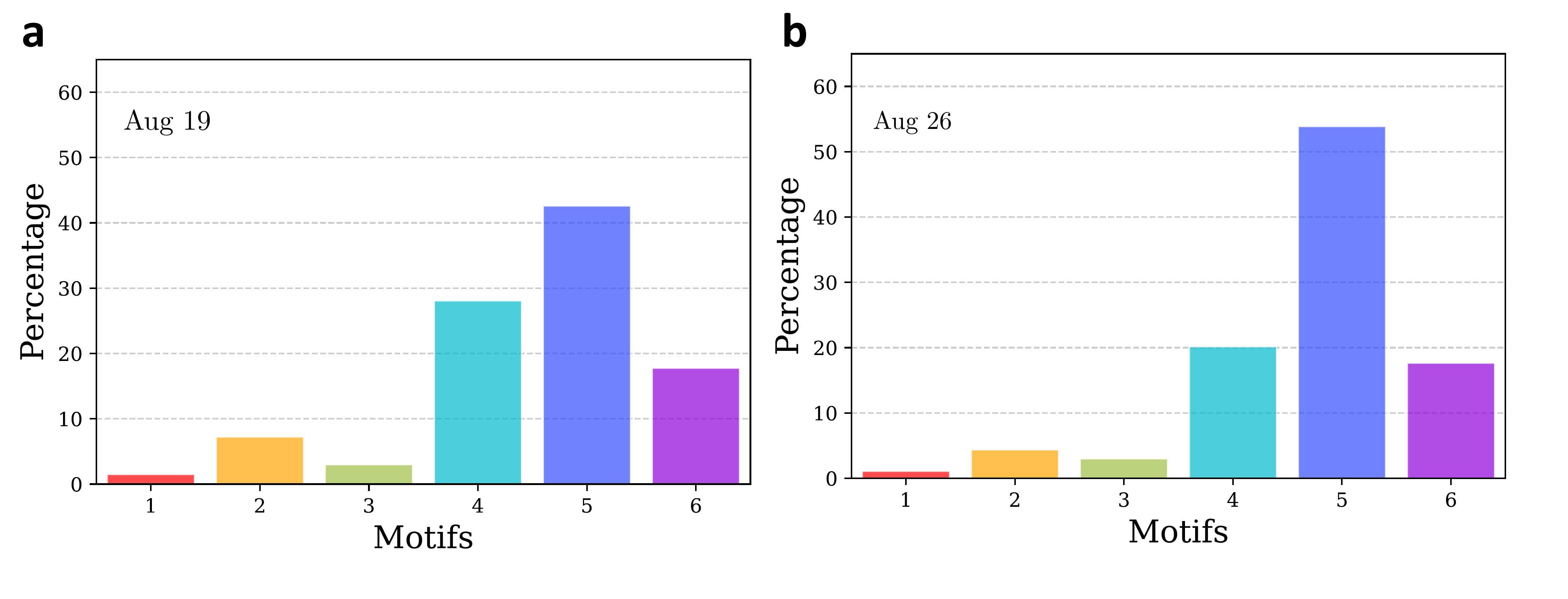}
\caption{Distribution of different motifs in terms of relative occurrence. Relative occurrence indicates the composition of the mobility network in terms of the six motif types. (a) for August 19, 2017 and (b) for August 26, 2017. During impact days (August 26), motifs show similar patterns of relative occurrence, indicating stability of order of relative occurrence even during disruptions.}
\label{fig:motif_conc_abs}
\end{figure}

Figure \ref{fig:motif_conc_abs} shows the motif distribution for August 19, 2017 (steady state) and August 26, 2017 (perturbed state), for the same day of the week. Evidently, for both stable and perturbed period, motifs demonstrate the following order of occurrence - \textit{$M_5$} > \textit{$M_4$} > \textit{$M_6$} > \textit{$M_2$} > \textit{$M_3$} > \textit{$M_1$}. Motifs 5 and 4 show the highest abundance and motifs 1 and 3 show the lowest frequency of occurrence. It is interesting to note that motifs 3 and 4, which have the same number of edges but differ in their structural orientation, show drastic differences in motif abundance. The same pattern is observed for motifs 5 and 6. 

\begin{figure}
\centering
\includegraphics[width=1\linewidth]{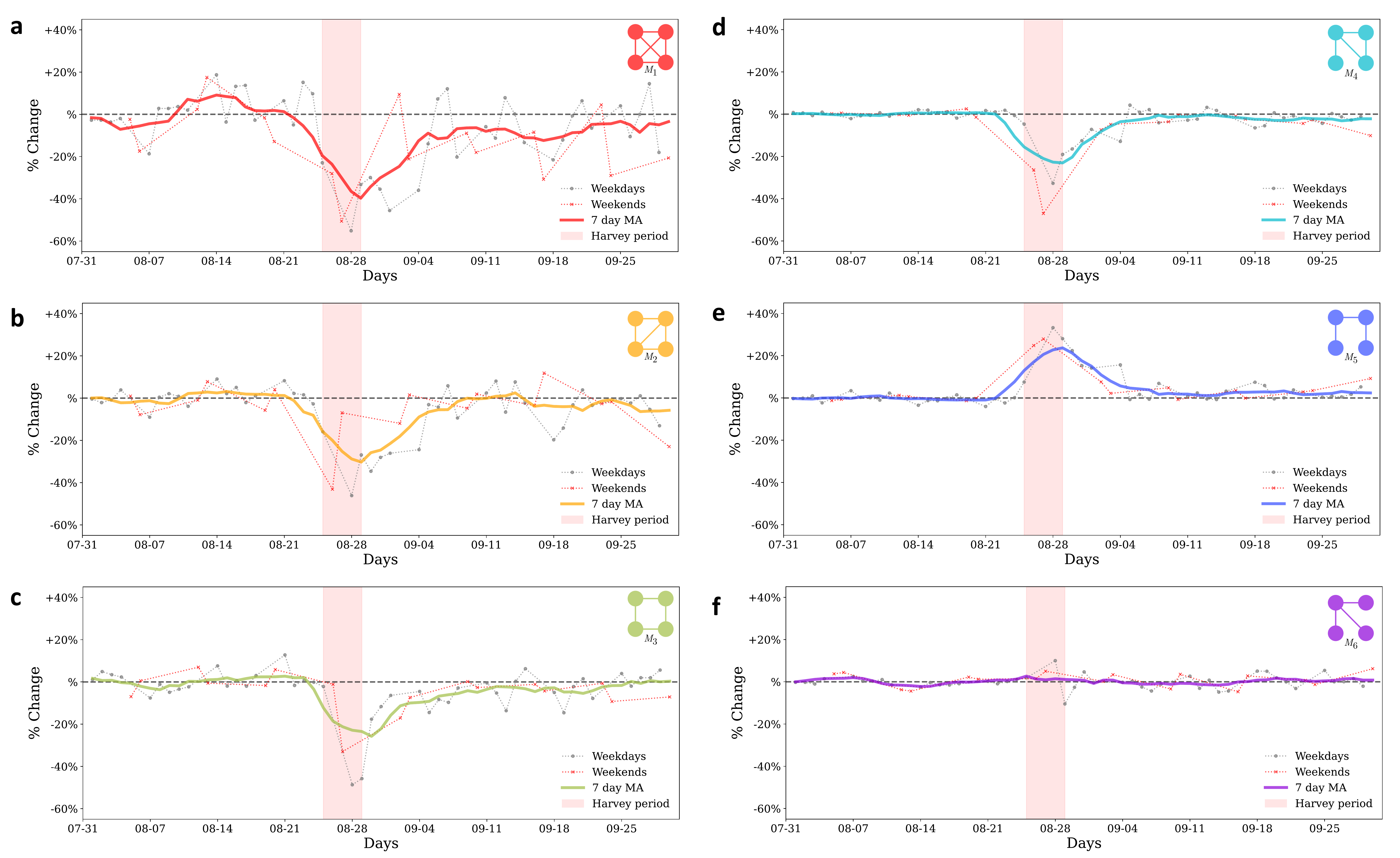}
\caption{Motif distribution change with respect to baseline for different motif types. (a) through(f) show the change in motif distribution for motifs 1 through 6, respectively. Percent change reflects the change with respect to baseline for the same day of the week during first two weeks of August. The densest motif, motif 1, shows the highest decrease during Harvey, while motif 6 remains unaffected. Motif 6 is the only motif that shows increase and may correspond to evacuation related travel as this motif structurally represents chain type of structure.}
\label{fig:motif_conc}
\end{figure}

We also computed the motif distribution change for each day as described in the methods and plot the time series obtained. Figure~\ref{fig:motif_conc} shows the motif distribution change of each of the motifs for weekdays, weekends, and 7-day moving average for all days. The changes in motif distribution indicate a change from the steady-state dynamics at the subgraph scale. The results suggest that motif distribution stays mostly stable during the steady state. This result suggests that motif distributions are stable in the absence of disturbances in human mobility networks. The changes in motif substructures in the network happened only on the day of the landfall of Harvey in Houston, August 25. The frequency of motifs 1-4 had a 50\% reduction in motif distribution for some of the days during the Harvey period. While most of the motifs showed a decrease in distribution, the frequency of motif 5 increased by 30\% some days (20\% of average increase), and Motif 6 remained fairly unaffected. For most of the motifs, maximum disruption occurred on August 26 and August 27. The observed changes happened on the first day of impact (hurricane landfall when people stayed sheltered in place or had already evacuated).

We observed that motifs 1 and 2, the most densely connected motifs) experienced the maximum impact showing a decrease of about 40\% and 30\%, respectively. Motif type 6, which models \textit{hub} and \textit{spoke} kind of movement showed close to no impact. This result suggests that travel patterns that involved census block groups as hubs (due to the presence of major points of interests) showed the highest robustness to flood-induced perturbations. Motif 5, which forms a chain, exhibits increased occurrence on the disaster days. This may correspond to relocation of residents to other areas or the relief and rescue efforts.  

For most of the motif types, near complete recovery of distributions to the steady state patterns can be observed in the week following September 4, but the abundance of motif 1 nodes still remained 10–15\% lower than the steady-state baseline for two more weeks, suggesting that dense movements had not fully recovered. A similar pattern can be weakly observed for motif 3 as well. Motifs 1 and 3 that are the least abundant (and most stable) during the steady state appear to be impacted for a longer duration compared to relatively more abundant (and less stable) motif types.   

\subsection*{Motif Persistence}
\begin{figure}
\centering
\includegraphics[width=0.85\linewidth]{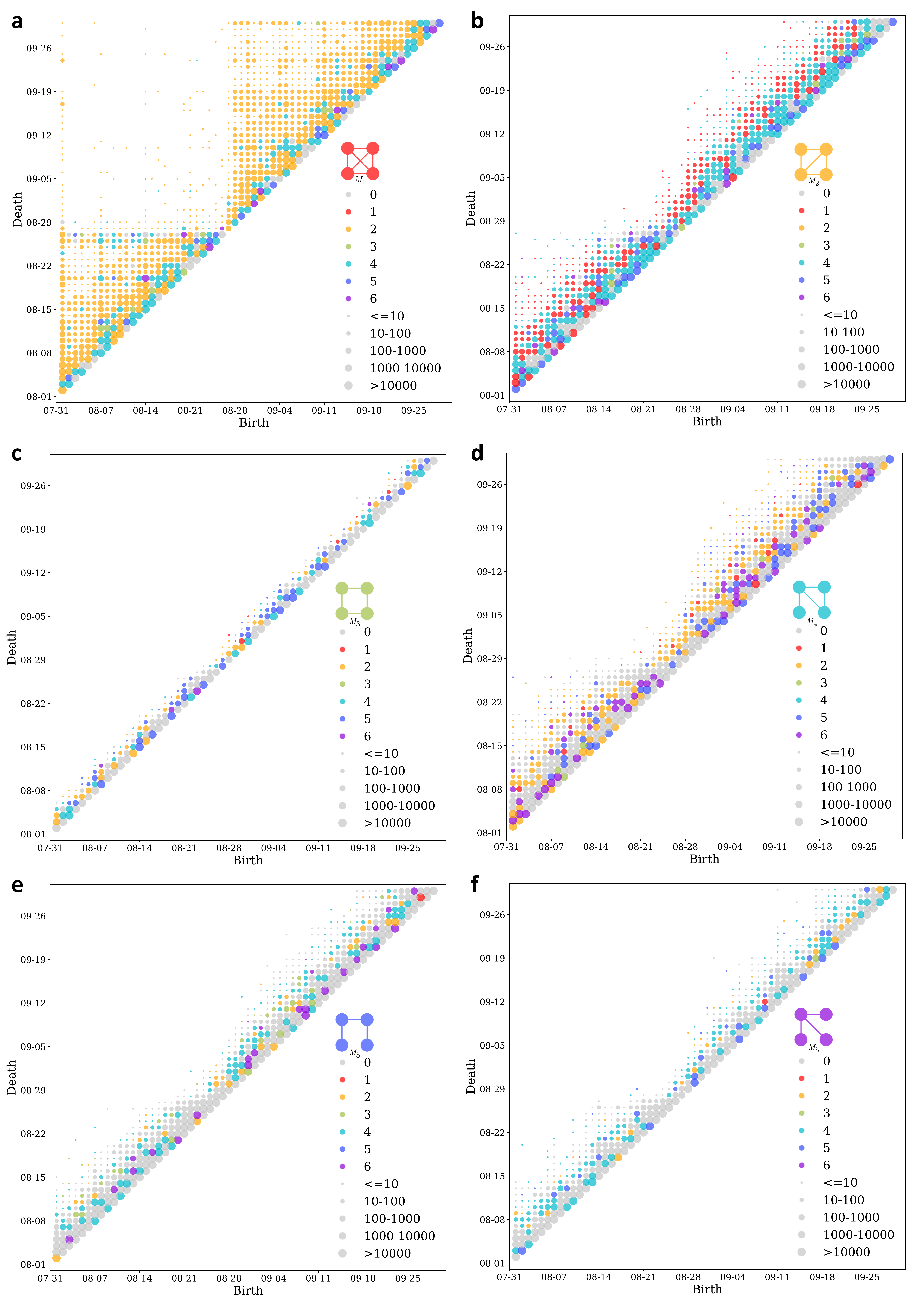}
\caption{Motif Persistence diagrams. (a) through (f) show the birth and death of re-sampled four-node pairs in the network for motifs 1 through 6, respectively. The markers on the plots represent specific conversions (birth, death event) based on the color and the number of four-node pairs undergoing that conversion (death) based on node size. Markers very close to the diagonal line imply that the motifs are very short-lived and temporally unstable.}
\label{fig:motif_PD}
\end{figure}

We identified that motif 1 experienced the highest impact due to Harvey flooding, and motif 6 remained fairly unaffected from the previous set of results. By observing the persistent diagrams for motifs, we analyzed whether these patterns also translate to temporal stability of motifs. Figure \ref{fig:motif_PD} shows the motif persistent diagrams for all motif types. Each marker in the figure represents a particular birth, death event for a set of node pairs. The colors represent the motif type that the four-node pairs convert to at the death event. If they become disconnected and do not fall in any of the motif types, they are represented by gray-colored marker. The size of the markers represent the number of four-node pairs showing same patterns for a birth, death event pair. For instance, if four-node pairs corresponded to Motif 1 on August 14 and converted to Motif 2 on September 10, they will be represented by orange marker of the smallest size in Figure~\ref{fig:motif_PD} at point August 14 to September 10. Markers closer to the diagonal imply that the time span between the birth and death of motif is short, or in other words, they exhibit a short life. Similarly, markers further from the diagonal mean longer life span.

It is clear from Figure~\ref{fig:motif_PD} that, at the death event of motif 1, most of the conversions happen to motif 2. Also, the markers are denser towards the diagonal line, which indicates that motifs in general are not very stable, and they have higher tendency to convert to other motif types, even in the steady-state period. Compared to other motif types, motif 1 appears to be temporally more stable as the markers also populate the areas away from the diagonal. A sparsely populated rectangular region in Figure~\ref{fig:motif_PD} for motif 1 shows that most of the motifs of this type that originated on or before August 27 covert to other motif types or become disconnected on August 28. This pattern is also weakly evident in motif 2 as well. Based on the spread of markers in the motif persistence plots, motif type 1 is the most stable type, followed by motif types 2 and 4, then motif types 5 and 6, and lastly motif type 3. It is also noteworthy that motifs that have one or more three-node cycles tend to have conversions to other motif types instead of being completely disconnected. For example, we find that (1) motif type 1 has higher tendency to convert to motif type 2, (2) motif type 2 has higher tendency to convert to motif type 1 or 4, and (3) motif type 4 converts equally to motif type 2 and being disconnected. Other motif types do not exhibit this behavior and have higher tendency to become disconnected and not fall in any of the six motif categories. This result suggests that motifs have an inherent tendency to preserve cycles. It is interesting to find that motif type 1, which shows the most decline in motif distribution, is the most stable motif during the steady-state period. Motif distribution of motif 6 is stable under perturbations but it is one of the least temporally stable motifs during the steady state. 

\subsection*{Motif Conversion}

From the motif persistence diagram, we can determine the temporal stability of motifs. To evaluate if the motifs are stable and to look for hidden mechanisms of motif conversions that help us examine network recovery, we analyzed motif conversion trends for each day. For example, if we look at conversion trends for motif type 1 for August 1, we calculate the percentage of node pairs that convert to different motif types or remain the same on August 2. We observe that these conversion patterns are similar for same days of the week. (Results can be found in the appendix.) Due to this similarity in motif conversion patterns for different days of the week, we plot conversion trends for all motif types including symbolic motif type 0 for different days of the week, separately.  

Figure \ref{fig:motif_conversions} of Appendix shows the results for motif conversions for all motif types for different days of the week. The days before and up to 3 weeks after the landfall of Harvey are marked separately in the figures to identify outlier days. Every motif, irrespective of the day of the week, shows a particular conversion pattern. For example, motif type 1 converts to same motif type for roughly 50\% of the time followed by motif type 2 and type 4, and it is less likely to convert to motif types 0, 3, 5 and 6. Although the percentages differ for different days of the week, the proportion of conversions and their patterns stay the same for all motif types. Motif types 1 and 2 are the most stable ones as they are more likely to retain their motif type with less conversion across days. Motif type 4 shows a similar conversion percentage to type 4, 5 and 0. Other motif types have higher tendency to convert to motif 0 or become disconnected. Motif type 3 is the least stable as it is less likely to retain its structure and mostly converts to motifs 1, 2, 4, 5 and 0, which is also evident from the motif persistence results. Although we see that these motifs show high temporal instability in terms of frequent conversions to other motifs types, their conversions follow distinct patterns. This indicates that even though the motifs appear to be temporally unstable, their conversion patterns are stable during nonimpact days.

Days corresponding to Harvey impact and flooding days (August 25 to August 29, 2017) and one day before the landfall (August 24) show different motif conversion trends than is evident in the rest of the data. The days up to one week after Harvey also show different conversion patterns. All motifs show higher conversions to motif type 0 on August 25, 26, and 28, which is also evident from the motif persistence plots. The days following landfall on August 25 till September 4, 2017 show higher conversions of other motif types to motif 1 and 2, indicating the tendency of sub-structures to recover through conversions to denser motif types that are also temporally more stable. After September 4, conversion patterns follow the usual trends indicating full or significant recovery after September 5, 2017.  This result shows that network recovery related to motif conversions happens within a week after the hurricane left Harris County. The network recovery date is based on motifs conversion patterns is similar to the one obtained from the global network measures (as discussed in future sections).

\subsection*{Motif Characteristics}

\begin{figure}
\centering
\includegraphics[width=0.9\linewidth]{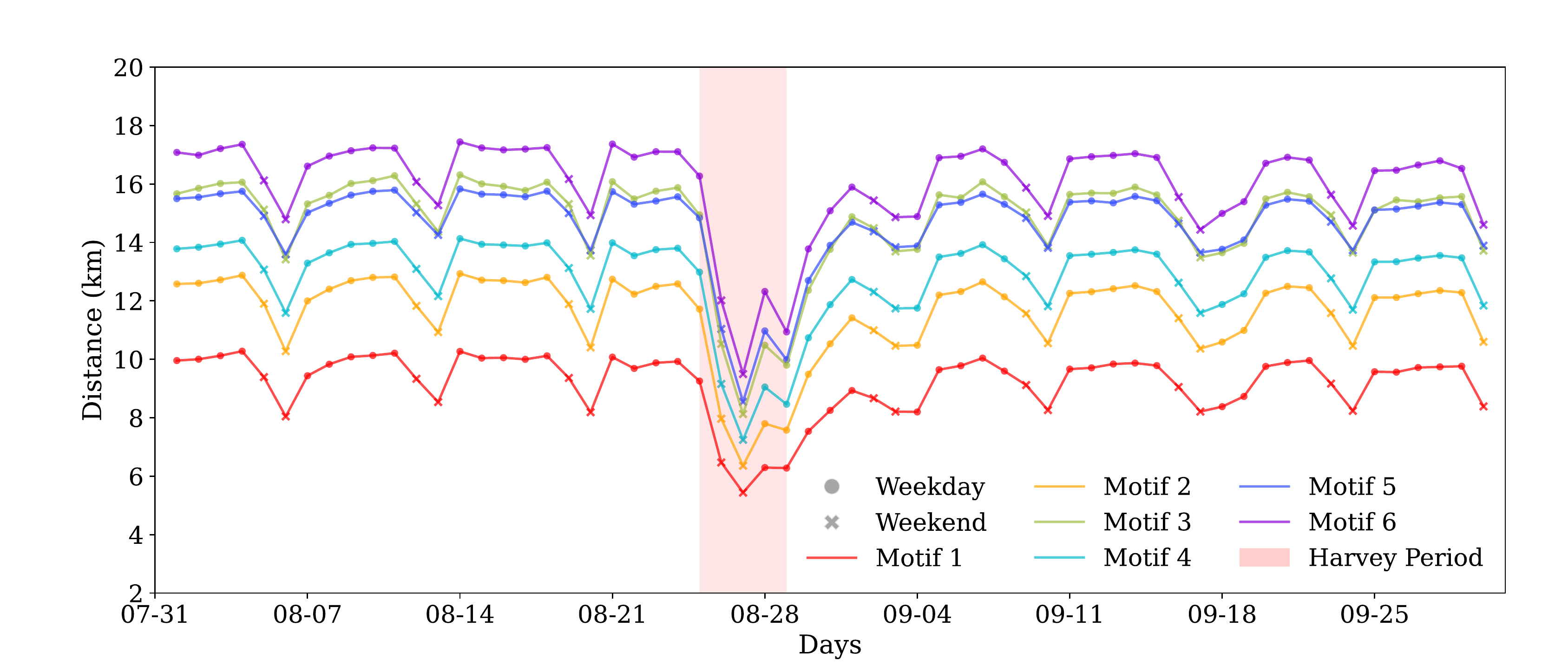}
\caption{Absolute values of median of the average motif distance. The figure represents the median average distance that the each of the motifs connect spatially. Denser motifs that have one or more three node cycles (motifs 1-3) tend to connect nearby places whereas less dense motifs connect longer distances on an average. During impact, all motifs show reduced distance indicating more localized travel.}
\label{fig:motif_dis_abs}
\end{figure}

\begin{figure}
\centering
\includegraphics[width=0.9\linewidth]{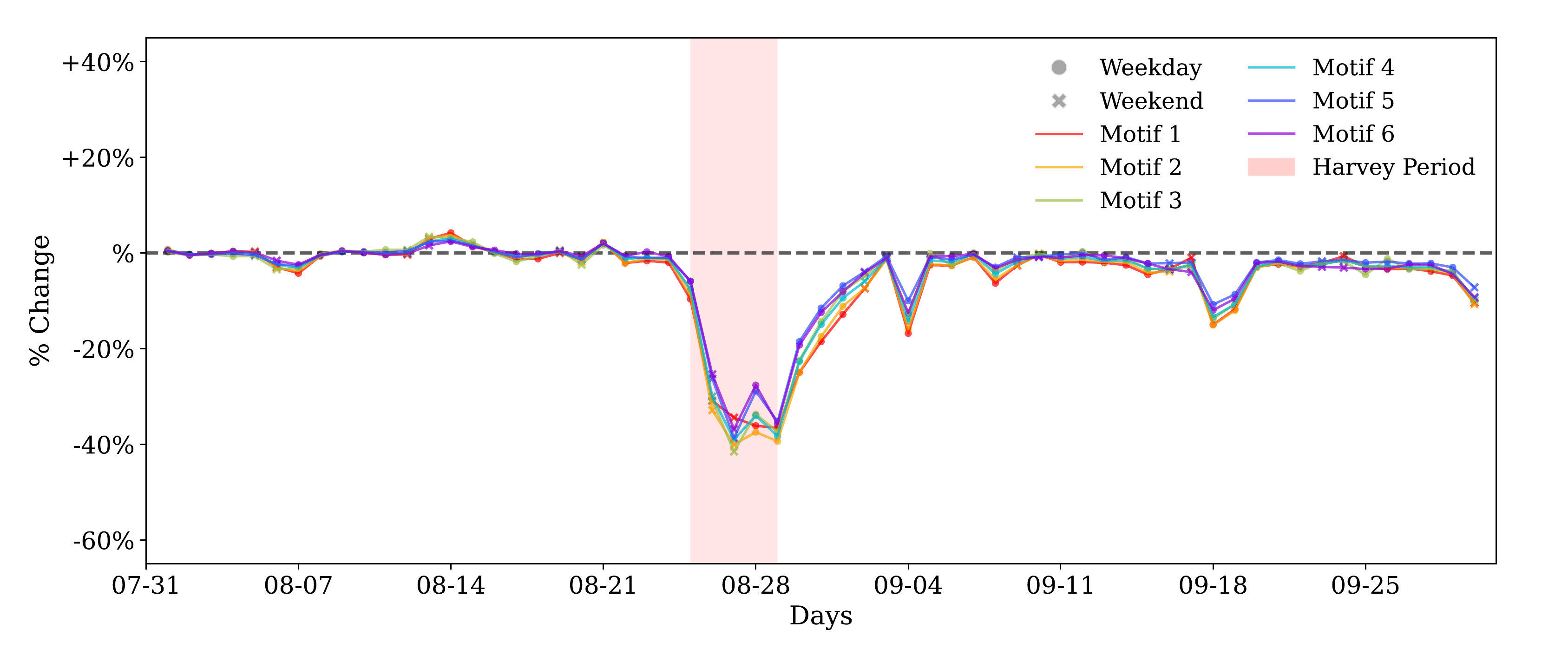}
\caption{Percentage change of median of the average motif distance compared to baseline. Extension to figure \ref{fig:motif_dis_abs}, this shows percent change in the distance that the motifs connect. The change in distance for all motifs seems to be similar, indicating that all motifs showed equal impact for this motif attribute.}
\label{fig:motif_dis_per}
\end{figure}

\begin{figure}
\centering
\includegraphics[width=0.9\linewidth]{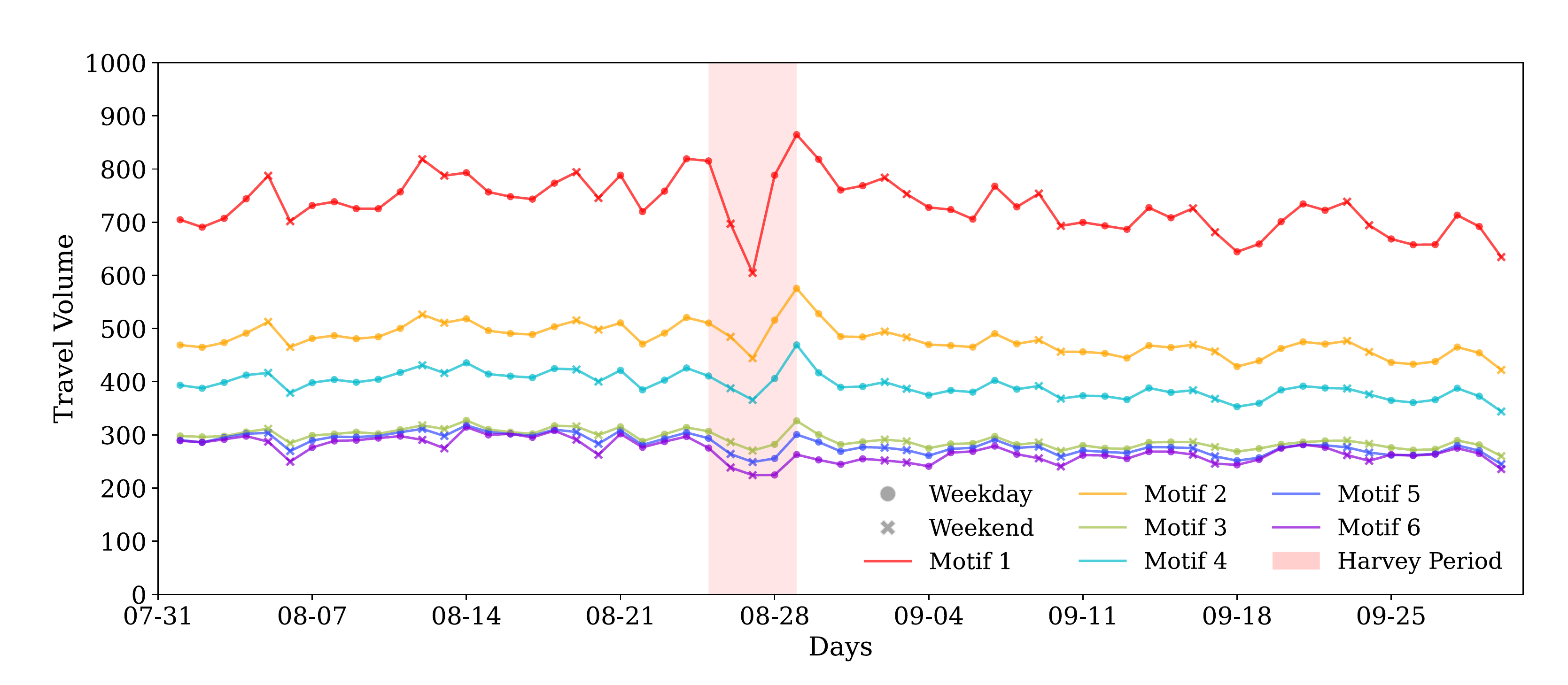}
\caption{Absolute values of median of the average motif travel volume. The figure represents the median of average travel volume that the each of the motifs represent. Denser motifs that have one or more three node cycles (motifs 1-3) tend to have higher travel volumes, whereas less dense motifs have very similar and low travel volume on an average.}
\label{fig:motif_vol_abs}
\end{figure}

\begin{figure}
\centering
\includegraphics[width=0.9\linewidth]{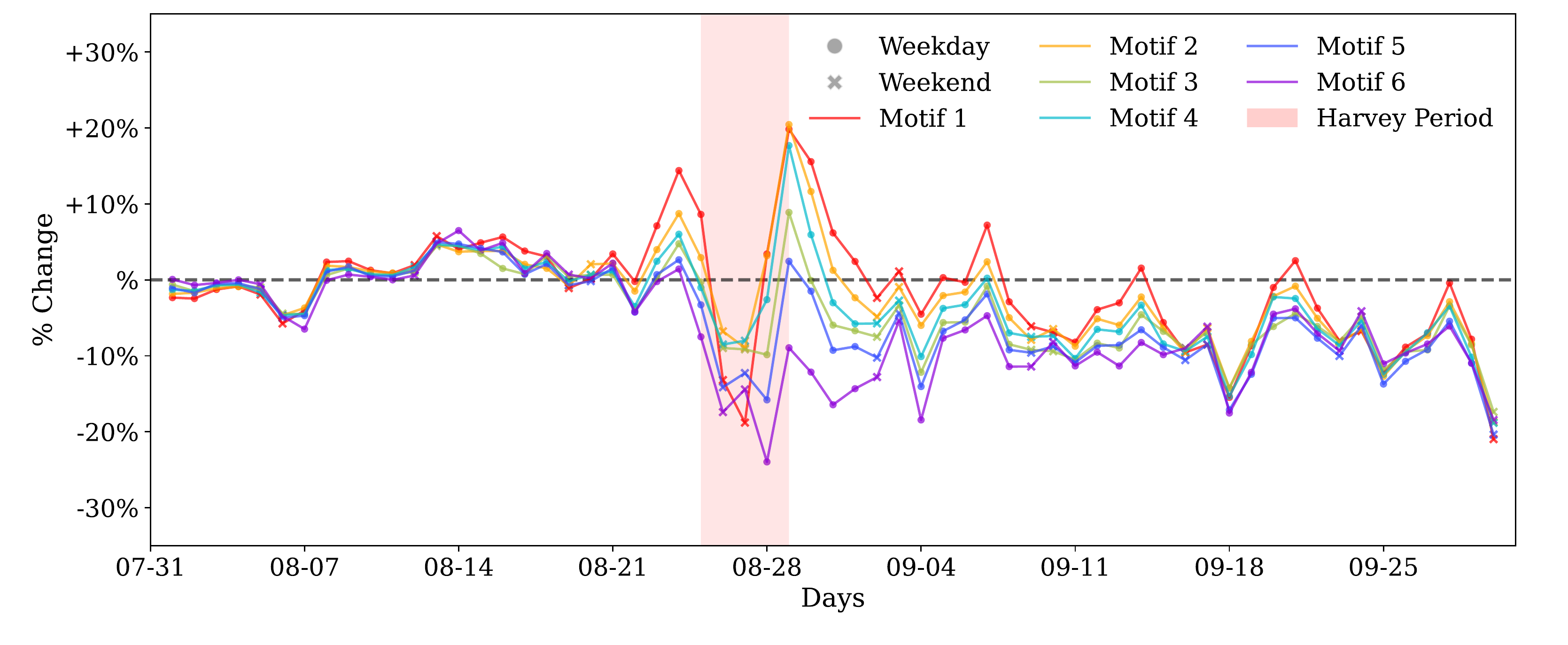}
\caption{Percentage change of median of the average motif travel volume compared to baseline. Extension to figure \ref{fig:motif_vol_abs} shown here is percent change in the travel volume that the motifs correspond to. There is a spike in travel volume for most of the motifs before and immediately after the impact duration which may correspond to evacuation based travel. The average travel volume decreases during impact for all of the motifs. Motif 6, corresponding to hub-spoke connection, shows the largest decrease. The average travel volume remains roughly 10\% lower than the baseline for the rest of the month for all motifs.}
\label{fig:motif_vol_per}
\end{figure}

So far we have focused on finding motif sub-structural changes in motifs. In the next step, we evaluate how motif properties, such as average travel volume and average distance, change with time. Figures \ref{fig:motif_dis_abs} and \ref{fig:motif_vol_abs} show the absolute values of median of the average motif distance and travel volume, respectively. Figures \ref{fig:motif_dis_per} and \ref{fig:motif_vol_per} show the percent change with respect to baseline of median of the average motif distance and travel volume respectively. From Figure \ref{fig:motif_dis_abs} it is clear that motif 6 connects the places that are far, followed by motifs 4 and 5. Motif 1 connects the nearest places, followed by motif type 2. During weekends, the distance reduces for all motif types, indicating closer trips, while the motif distribution remain the same. Moreover, based on the results from the previous section, we can say that temporally stable motifs are formed by trips with shorter distances, and least-affected motif based on concentration change connects far apart places. The results of the percent change in the motif distance metric indicates that all motif types got affected very similarly as the curves are plotted very close to each other. There is a 40\% decrease in distance for all of the motifs during the perturbed period. This disruption happens immediately after landfall of Harvey and recovers within a week. There appears to be no permanent impact in the motif distance characteristic beyond one week from the landfall. 

Travel volume shows contrasting but intuitive trends as compared to motif distance metric. Motif 1 corresponds to the highest travel volume, and motifs 6, 5, and 3 have the lowest. The percent change in the travel volume with respect to the baseline shows an increase in travel volume a day before the landfall of Harvey and on the last day of impact period, August 29, 2017. An increase in movement before to landfall suggests pre-disaster preparedness efforts and the increase on the last day of the perturbed period suggests response efforts after Harvey reduced in intensity. It is interesting to note that motif structures did not exhibit any structural changes before landfall, but travel volume motif characteristic is able to capture these movements. While most of the structural changes in the motifs recover within a week of landfall, motif travel volume seems to show a short-term impact with around 10\% less travel volume sustained until the end of September. 

\subsection*{Global Network Properties}

The previous results focused on the sub-structure characteristics and the impact of perturbations and recovery at the motif level. In this section, we present results related to the global network properties, such as average network diameter, modularity, density, average degree, diameter, and giant component size to evaluate impact of Harvey on at global network scale for comparison with the impact extent and recovery at motif level.

While giant component size is one of the metrics researchers have used to evaluate network resilience, in our case, the network is more densely connected, so the giant component size remains the same as the total number of nodes (500) throughout the analysis period. In the case of human mobility network, giant component size is not able to reflect changes due to impact of Harvey, indicating one of its drawbacks for dense networks for characterizing resiliency. 

The network diameter which increased by 60\% during the Harvey period and returned to the pre-disaster value on August 31. This change indicates earlier indication of recovery at this global network property. Network modularity that evaluates the level of clustering in the network shows similar trends to that observed for micro-structures, that is, motifs. We observed increased modularity and decreased average node degree and density in the network during the Harvey period. The recovery takes place within one week after Harvey dissipated (the same recovery duration that we observed in motif conversion and attribute analysis). This result indicates that properties such as modularity, average node degree, and network density are better able to reflect micro-structure dynamics for network resilience characterization. However, these global network measures could capture only some aspects of network sub-structure impacts and recovery dominated by the characteristics of abundant (and less stable) motifs. As shown earlier, the impact of flooding on the frequency of less frequent (but more stable) motifs is more extensive, and its recovery takes longer than what global network properties indicate.

\iffalse
\begin{figure}
\centering
\includegraphics[width=0.7\linewidth]{Figures/Network_Diameter.png}
\caption{Network diameter.}
\label{fig:net_dia}
\end{figure}

\begin{figure}
\centering
\includegraphics[width=0.7\linewidth]{Figures/Network_Modularity.png}
\caption{Network modularity.}
\label{fig:net_mod}
\end{figure}
\fi

\begin{figure}
\centering
\includegraphics[width=0.9\linewidth]{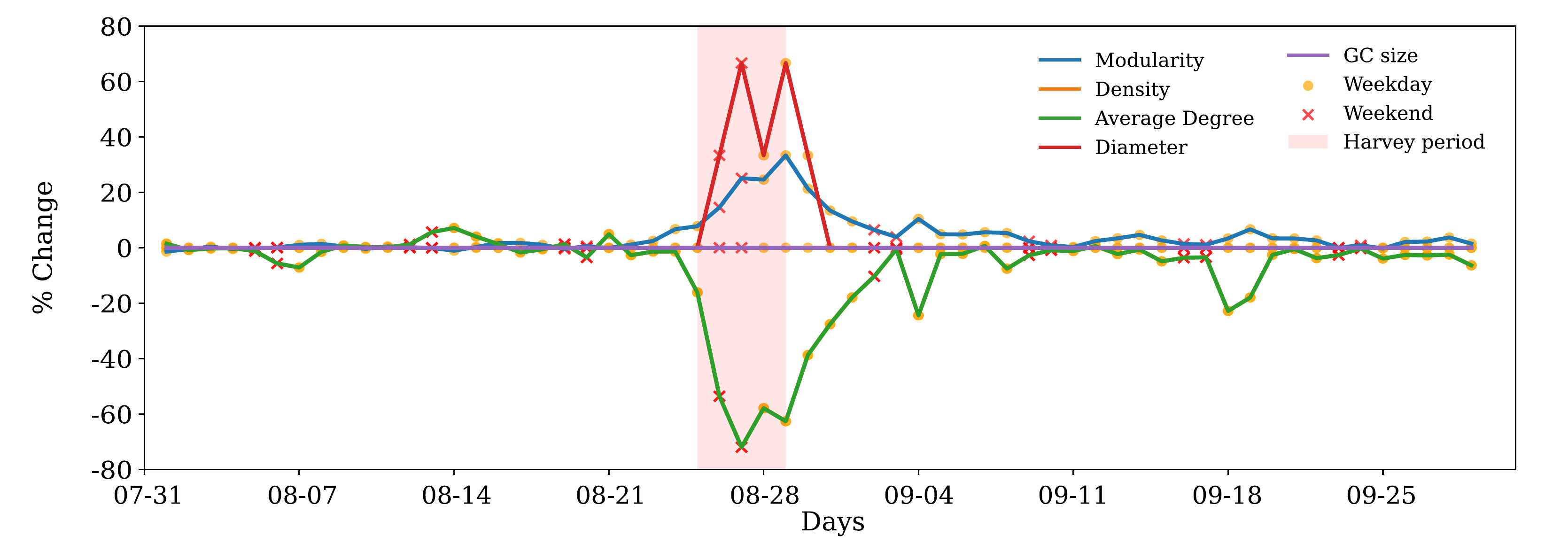}
\caption{Global network measures. Impact of Harvey on global network measures, such as giant component size, modularity, network density, average node degree, and diameter is shown. Plot of network density is not visible as it exactly overlaps with average degree since there are 500 nodes in each time step in the giant component.}
\label{fig:net_dia}
\end{figure}

\section*{Discussion}

The main idea of this study is that, in studying resilience in temporal human networks, solely relying on global network measures would be inadequate; latent sub-structural network mechanisms need to be examined to determine the extent of impact and recovery of these networks during perturbations such as urban flooding. In this study, we utilized high-resolution aggregated location-based data to construct the temporal human mobility networks in Harris County, Texas (Houston metropolitan area) in the context of the 2017 Hurricane Harvey. Using the constructed temporal network models, we examined various characteristics, such as motif distribution, motif persistence and temporal stability, and motif attributes to reveal latent sub-structural mechanisms related to the resilience of human mobility networks during disaster-induced perturbations.

The analysis of network micro-structures suggests that motifs in mobility networks, even during steady states, do not exhibit high temporal stability. In general, frequently occurring motifs (e.g., motif 5) are temporally one of the least stable motifs, and structurally dense motifs (e.g., motif 1) are more stable temporally but are not abundant in the network. The reason for this dichotomy of abundance and temporal stability in motifs of human mobility networks is that that motifs connecting shorter distances have high movement volume, and those connecting farther places have lower travel volume (consistent with the main idea of gravity model but at the motif level ~\cite{Hong2019}). Hence, low volume links are more prone to falling below the significant number of trips threshold (and thus, become temporally unstable). This dichotomy of abundance and temporal stability in motifs of human mobility networks influences the way perturbation impacts and recovery are manifested in the global network measures. The abundant (and less temporally stable motifs) experience relatively lower motif frequency change due to flood perturbations and faster recovery. The indication of network impact and recovery measure by global network metrics is similar to the impact extent and recovery duration of abundant motifs. Global network measures might indicate a network has recovered, while less abundant motif types are still recovering in terms of their frequency in the overall network.

During perturbation impacts, we observe that motifs exhibit significant changes in their structure and attributes. In the perturbed state, motifs have different distributions; most show a decrease in motif count. The hub-and-spoke type of motif (motif 6) remains unchanged; a chain type of motif (motif 5) shows an increase. Distribution of motif 1 and average travel volume of motifs show extended impact during flooding, implying sustained perturbation in these motifs while other global network measure show full network recovery. The average distance that the motifs connect decreases by up to 40\% during flooding indicating that motifs corresponding to more short distance travel took longer to recover. This result implies that densely connected motifs corresponding to short-distance trips experience a greater sustained trip volume reduction, and thus take longer to recover from urban flooding.  

The analysis of change in motif travel volume also indicates a 15-20\% increase in movement volume a day before and on the last day of impact duration, suggesting preparedness of the residents for Hurricane Harvey. This pattern of preparedness behavior could not be captured through global network properties and structural attributes of motifs in human mobility networks. 

Motifs also exhibit different conversion patterns during flood-induced perturbations compared with the steady state conversion patterns. On the day of the landfall, most of the motifs break down into disconnected structures, but immediately after impact, motifs show high conversions to denser motifs (motif 1 and 2), indicating tendency of motifs to convert to temporally stable motifs as they recover from perturbations. As motifs recover, the conversion patterns return to the steady-state conversion patterns.    

One of the important and major findings of this study is that commonly used global network properties, such as network diameter and giant component size to evaluate resilience of a network, do not fully capture the underlying failure mechanisms in temporal networks. For example, in the case of our human mobility network recovery, network diameter indicated recovery within two days after the disaster, but the sub-structure motifs in general took one week to indicate recovery, and some characteristics, such as movement volume showed recovery was not achieved till the end of September 2017. Although other global network measures, such as modularity, average degree, and network density, show disruption duration up to one week after Harvey, they do not provide any insight regarding the slower recovery of less abundant motifs that sustained for a longer period. Also, we are able to capture pre-disaster preparedness behavior-based movements through motifs while these preparedness patterns are not apparent in the global network properties. The inadequacy of global network measures for resilience characterization in temporal networks is due to the fact that we observe unified structural changes when we look at global properties, but the analysis of substructure level changes reveal how changes in motif distributions, their conversion pattern, and attribute variation all build up to global failure. Therefore, to understand the resilience of temporal human mobility networks (and other temporal networks), it is essential to examine the microstructures and their perturbation impacts and recovery. From a practical perspective, the results can provide disaster managers, public officials, and transportation planners with insights to better evaluate impacts and monitor recovery in affected communities based on the patterns of impact and recovery in human mobility networks at both substructure and global network levels.   

Our study has certain limitations. For the sake of computational efficiency, we took four-node undirected motifs, but the work can be expanded to directed motifs to better explain the flow of population, especially for the hub-and-spoke motif (motif 6). In addition, our study does not incorporate node attributes, such as presence of points of interest, spatial area characteristics (e.g., residential zone, business zone, and school), and population density. Studies generalizing motifs by considering directed motifs and incorporating node attributes may be able to provide additional insights into substructure behavior and may give some universal laws governing network microstructures associated with resilience.

\section*{Acknowledgments}
The authors would like to acknowledge the funding support from the National Science Foundation CRISP 2.0 Award under grant number 1832662. The authors would also like to acknowledge Streetlight Data for providing the mobility data. Any opinions, findings, conclusions, or recommendations expressed in this research are those of the authors and do not necessarily reflect the view of the funding agencies.
%Bibliography

\bibliographystyle{unsrt}  
\bibliography{arxiv}

\newpage

\appendix
\section*{Appendix}
\renewcommand{\thefigure}{A\arabic{figure}}
\setcounter{figure}{0}

\begin{figure}[htp]
\centering
\includegraphics[width=0.99\linewidth, angle=90]{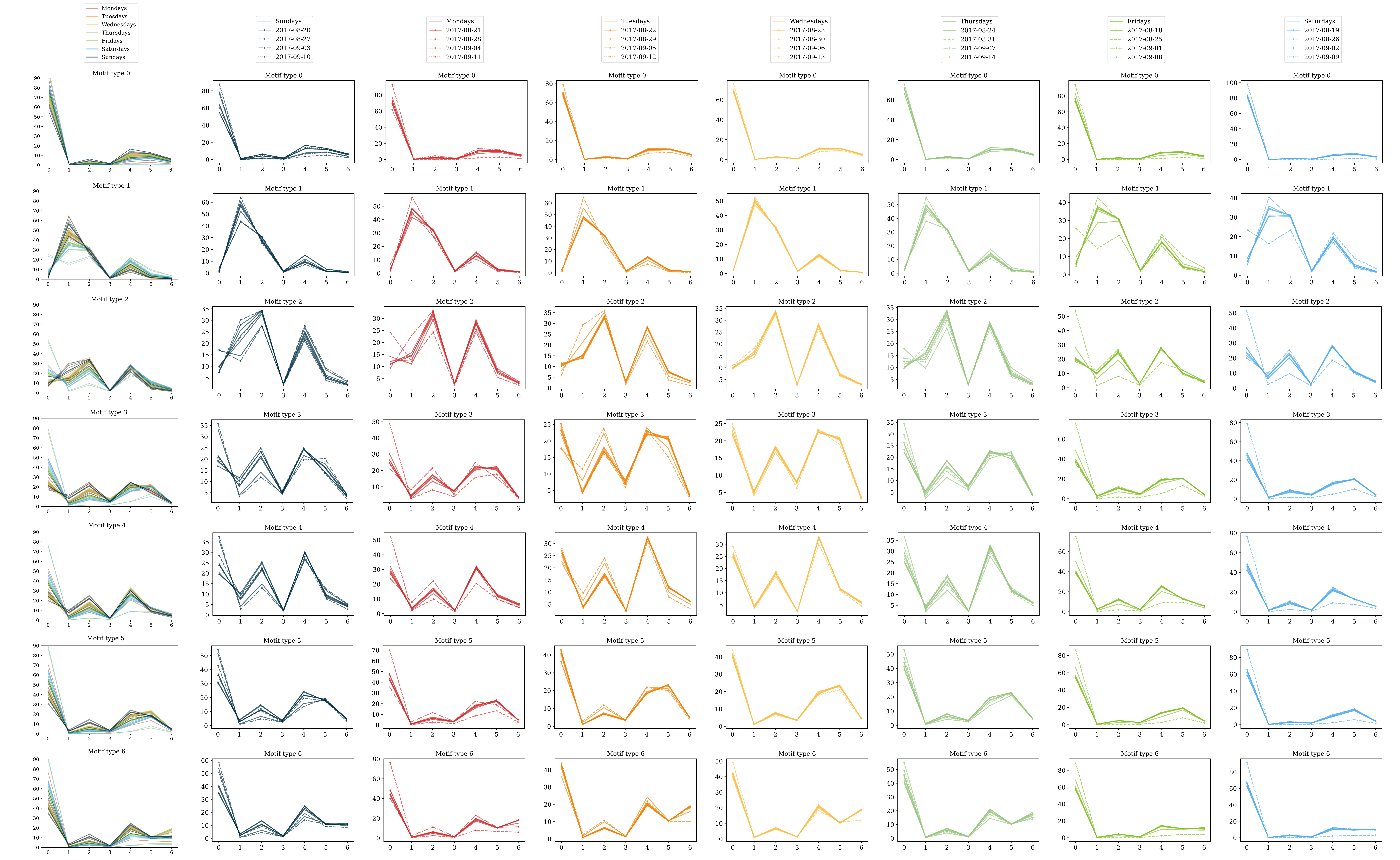}
\caption{Motif conversion trends for different days of the week: first column shows the plots of all the days, with each day of the week color coded separately. Other columns show motif conversions for different days of the week and some days before, during, and immediately after the impact are highlighted for better interpretation of the results.}
\label{fig:motif_conversions}
\end{figure}

\end{document}